\documentclass[aps,prl,floats,twocolumn,letterpaper,amssymb,
showpacs,floatfix]{revtex4}
\usepackage{graphicx}

\newcommand{\nicefrac}[2]{\leavevmode\kern.1em
    \raise.5ex\hbox{\the\scriptfont0 #1}\kern-.1em
    /\kern-.15em\lower.25ex\hbox{\the\scriptfont0 #2}}
\begin{document}
\title{Capture-zone scaling in island nucleation: phenomenological theory of an example of universal fluctuation behavior}
\author{Alberto Pimpinelli$^{1,2,}$}
\email[]{alpimpin@univ-bpclermont.fr .......     **einstein@umd.edu}
\author{T. L. Einstein$^{2,}$$^*$$^*$}
\affiliation{$^1$LASMEA, UMR 6602 CNRS/Universit\'e Blaise Pascal -- Clermont 2, 
F-63177 Aubi\`ere cedex, France
\\ $^2$Department of Physics, University of Maryland, College Park, Maryland 20742-4111 USA}

\date{\today}

\begin{abstract}
In studies of island nucleation and growth, the distribution of capture zones, essentially proximity cells, can give more insight than island-size distributions.  In contrast to the complicated expressions, ad hoc or derived from rate equations, usually used, we find the capture-zone distribution can be described by a simple expression generalizing the Wigner surmise from random matrix theory that accounts for the distribution of spacings in a host of fluctuation phenomena.  Furthermore, its single adjustable parameter can be simply related to the critical nucleus of growth models and the substrate dimensionality.  We compare with extensive published kinetic Monte Carlo data and limited experimental data.  A phenomenological theory sheds light on the result.
\end{abstract}


\maketitle

In the active field of statistical mechanics applied to materials, an important unsettled problem in morphological evolution during epitaxial thin film growth \cite{evans2} 
is the characterization of the
statistical properties of nucleating islands. In particular,
for over a decade the universal scaling shape of
the island-size distribution (ISD) has been investigated numerically with
kinetic Monte Carlo (kMC) simulations, but analytical evaluation has proved elusive.  Only 
rate equations \cite{amfam,amarp,amars} or complicated (often implicit) expressions \cite{bm1996,mr2000} have been proposed. 
The ISD is an important tool for experimentalists, since simulations have shown
it to be a unique function of
the size $i$ of the critical nucleus (see below), a quantity that describes the largest 
unstable cluster.

A decade ago Blackman and Mulheran \cite{bm1996,muhl}
proposed subordinating
the ISD to the distribution of areas of Voronoi polygons (proximity cells) built around the
nucleation centers. Once an
island is nucleated, it efficiently captures most of the adatoms diffusing
within the capture zone (CZ), a region roughly coinciding with the island's Voronoi polygon. 
This breakthrough, which invites analogies to distributions of quantum dots \cite{Fanf} and to foams \cite{WKW}, led to several 
part-numerical, part-analytic investigations \cite{evans2,amfam,amarp,amars,bm1996,mr2000} that allowed 
prediction of the ISD for point
islands with good accuracy, at the price of performing extensive kMC
simulations or of solving a
system of several coupled, non-linear rate equations, which is computationally as
taxing as kMC. For this reason,
an empirical functional form, proposed in Ref.~\cite{amfam}, which fits kMC results well, 
is still widely used to analyze data.

In this Letter, we propose a different approach. We show that 
the generalized Wigner surmise (GWS) distribution, a
class of probability distribution
functions rooted in random matrix theory
(RMT) \cite{mehta,guhr}, yields an
excellent quantitative description of the CZ size distributions for all values
of the critical nucleus size $i$ in published
simulations. Thus, this relatively mature subject is vitalized and broadened 
by linkage to universal aspects of fluctuations. RMT experts will find remarkable 
that the signature exponent has atomistic meaning in these non-equilibrium systems.  
A phenomenological
argument suggests the physical origins of the GWS here. 
\begin{figure}[b]
\includegraphics{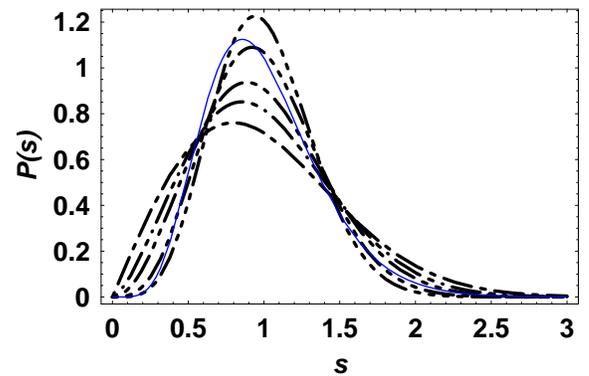}
\caption {[Color online] Plots of the GWS of Eq.~(\ref{eq:Ps}) $P_n(s)$, $n$=1,2,3,4, of relevance in this paper, indicated by long dashes alternating with $n$ short dashes; also $P_{3/2}(s)$, indicated by long dash, short dash, dot.  The thin solid [blue] curve shows the Gamma distribution $\Pi_7(s)$, discussed later. 
}
{\label{f:WigGam}}
\end{figure} 

RMT \cite{mehta,guhr}
has been successfully applied as a
phenomenological description of statistical fluctuations in a large variety of
physical systems, such as highly
excited energy levels of atomic nuclei, quantum chaos \cite{Haake}, 
cross-correlations in financial data \cite{GuhrStanley},
stepped crystal surfaces\cite{bart}, and even times between buses in Cuernavaca \cite{bus} and distances between parked cars \cite{Abul}! The last example is analogous to our study in that the RMT-derived formula accounts for the data notably better than the culmination of years of problem-specific theories. 

RMT applies only to matrices with
special symmetries, which constrains
the applications to physical systems that somehow reflect these symmetry properties. The Wigner surmise $P_{\beta}(s)$ 

\begin{equation}
P_{\beta}(s)= a_\beta s^\beta \exp(-b_\beta s^2)
\label{eq:Ps}
\end{equation}
(cf.\ Fig.~\ref{f:WigGam}) provides a simple, excellent approximation for the distribution of spacings for such cases \cite{mehta,guhr}.  
\noindent Here $s$ is the fluctuating variable divided by its mean, $\beta$ is the sole WS parameter \cite{varrho}, and  $a_\beta$ and $b_\beta$
are fixed by the normalization
and the unit-mean conditions \cite{ab}.

\begin{figure}[t]
\scalebox{0.64}{\includegraphics[10,0][350,230]{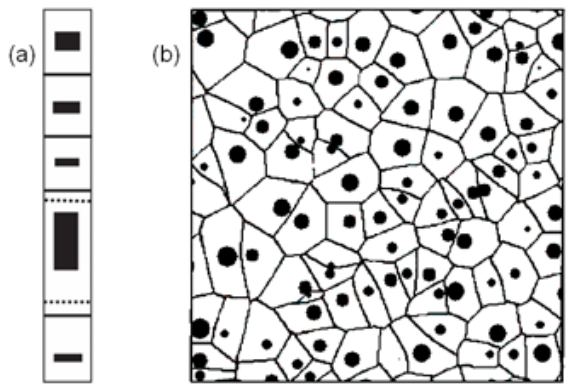}}
\caption { a) Schematic for 1D (vertical). Black rectangles correspond to 1D islands. Horizontal lines mark the midpoints between the {\it edges} of two neighboring islands, with the capture zones (CZ) defined as the resulting proximity cells.  An alternative definition, implicit in the point-island approximation, uses the midpoint between the {\it centers} of islands, indicated by dashed lines and leading to Voronoi ``cells".  For islands [nearly] the same size, the two [nearly] coincide. b) 2D illustration of the islands (approximated as circular) and the Voronoi polygons that bound their CZ, from Ref.~\cite{muhl}.
}
{\label{f:isdcz}}
\end{figure} 

Standard RMT \cite{mehta,guhr,Haake} fixes attention on the values 
1, 2 and 4  of $\beta$, corresponding to orthogonal, unitary, and
symplectic matrices, respectively.  
The GWS posits that Eq.~(\ref{eq:Ps}) has physical relevance for 
{\it general} non-negative $\beta$ \cite{einst1}.
We show here that {\it the CZ distribution
is excellently described by the GWS with} $(\nicefrac{\footnotesize 2}{\it d})(i\! +\! 1)$, where
$d$=1,2 is the spatial dimension (see Fig.~\ref{f:isdcz}).  The GWS also describes the distribution of terrace widths
on stepped surfaces \cite{bart,einst1}, where the step-repulsion strength determines $\beta$.

The explicit dependence of $\beta$ on dimension is a novel feature of this study.  Most other applications of RMT are either essentially one-dimensional or insensitive to $d$.  While $P_1(s)$ describes well the nearest-neighbor spacings between randomly-distributed points on a plane \cite{Haake}, GWS fits of Voronoi tessellations of such points are not particularly good \cite{lecaer,dw}. 
For island nucleation, subtle correlations between nucleation centers lead to a distribution of tesselation cells (Fig.~\ref{f:isdcz}) described by the GWS.

\begin{figure}[t]
\scalebox{0.8}{\includegraphics[2,0][290,200]{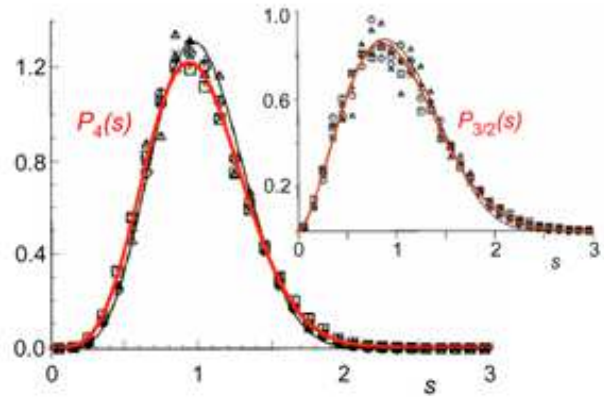}}
\caption {[Color online] CZ size distributions for critical nucleus size $i\! =\! 1$ and $d\! =\! 1$.  Symbols are from Fig.~12 of Ref.~\cite{bm1996}, for various pairs of values of ($D/F$ in units of 10$^5$, coverage in monolayers): $\boxdot$ (5, 0.11), $\odot$ (5, 1.19), $\bigtriangleup$ (5, 12.65), $\times$ (50, 0.11), $+$ (50, 1.19).    
The thin curve is the theory prediction of Ref.~\cite{bm1996}.  The thick [red] curve is the simpler $P_4(s)$.  The inset shows similar results for distribution of {\it gaps} between point islands from Fig.~11 of Ref.~\cite{bm1996}, with the added thick [red] curve giving $P_{3/2}(s)$; its self-convolution is about $P_4(s)$, the ISD, as discussed near the end.
}
\label{f:bm}
\end{figure} 

Island nucleation is pictured as atoms deposited on a substrate (at rate
$F$) and then diffusing on the surface at diffusion rate $D$ (most properties depending only on $D/F$ \cite{evans2}).
When adatoms meet, they form bonds, whose lifetime depends on temperature $T$. At
low enough $T$, bonding is
virtually irreversible, so that an adatom pair is a stable---and immobile---island, 
which grows only by
capturing other adatoms.  A single adatom is then called a {\it critical
nucleus}; equivalently, the
critical nucleus size is $i=1$ at low $T$. At higher $T$ a single bond
will be broken before other adatoms
can be captured, and the critical nucleus will be a larger cluster, whose size
will depend on the surface lattice
symmetry, generally $i=2$ or $3$ on a (111) or (100) surface, respectively \cite{amfam,zang}.

We first test our approach on data computed by Blackman and Mulheran \cite{bm1996}
with kMC simulations of the nucleation of point islands
along a one-dimensional (1D) substrate (cf.\ Fig.~\ref{f:isdcz}a). Since $i$=1 there, we predict that the CZ size
distribution is a GWS function
with $\beta$=4. Fig.~\ref{f:bm} shows the results of their
simulations, along with fits with the Wigner surmise. Clearly
$P_{4}(s)$ yields an excellent fit to the numerical results, better than the thin solid line, 
the result \cite{bm1996} of a
statistical numerical calculation replacing the solution of a complicated
integro-differential equation.  Thus, our expression is both more accurate and simpler than their theoretical  result.

\begin{figure}[t]
\scalebox{0.84}{\includegraphics[0,0][216,250]{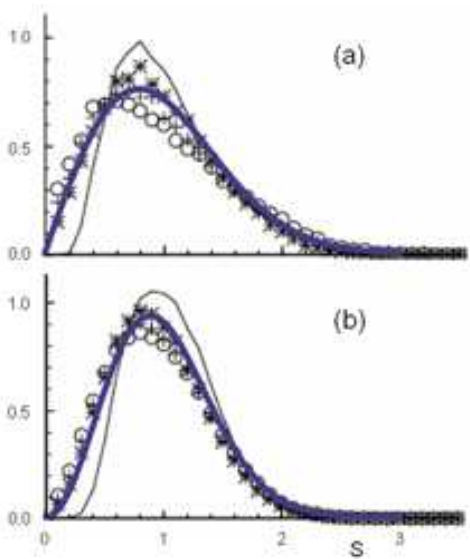}}
\caption {[Color online]  (a) Symbols are numerical data from Fig.~2b of Ref.~\cite{mr2000},
giving the CZ size distribution for
nucleation of islands with $i\! =\! 0$ in 2D. The thick (blue) curve is  $P_1(s)$.  (b) Same as panel (a), but symbols for $i=1$ from Fig.~2d of Ref.~\cite{mr2000}, and the thick curve 
is $P_2(s)$. In both panels the thin curve is the theory of
Ref.~\cite{mr2000}.
}
{\label{f:mr}}
\end{figure}
Two-dimensional (2D) deposition, diffusion, and aggregation
models have been
extensively treated by many authors. 
Mulheran and Blackman \cite{muhl} report kMC simulations of growth of fractal islands ($i\! =\! 1$)
and circular islands
($i\! =\! 1$ to 3). 
For the circular islands we find very good agreement between the data and the GWS using $\beta\! =\! (\nicefrac{2}{2})(i\! +\! 1)$ \cite{PE}, with the trend for increasing $i$ well reproduced. Even better agreement is found between the GWS $P_{i+1}(s)$ and Mulheran and Robbie's
\cite{mr2000} more recent
kMC simulations of nucleation and growth of circular
islands for $i\! =\! 0$ and 1,
as shown in  Fig.~\ref{f:mr}a and \ref{f:mr}b, again superior to their numerical-analytical theory \cite{mr2000}.
Popescu et al.~\cite{paf} also report extensive kMC simulation data
of irreversible nucleation ($i=1$) of point, compact, and fractal islands, but do not compute CZ size
distributions. Their rate-equation approach was designed to describe island sizes and capture numbers,
so should not, and does not \cite{PE}, describe the CZ distribution well.

To understand why the CZ distribution is well described by $P_\beta(s)$ with $\beta \! =\! (\nicefrac{\footnotesize 2}{\it d})(i\! +\! 1)$,
we offer a phenomenological model.
We draw on our recent demonstration \cite{pimpi1} that the GWS appears in the
context of RMT as the mean-field
solution of Dyson's Brownian motion model \cite{guhr,Haake}, based on a Coulomb gas of logarithmically interacting particles \cite{notlog} in a 1D quadratic
potential well.  We argue that the CZ size distribution can be extracted from a Langevin
equation for a fluctuating CZ size
in a confining
potential well created by two competing
effects:    1) The effective confining potential well 
should increase for small-size CZ: nucleation of a
new, small island causes a CZ of finite (and not greatly different from the mean) size to appear (cf.~Fig.~19 of \cite{evans1}), so that a large
force must prevent fluctuations of
the CZ size towards vanishing small values  2) The neighboring CZ's also prevent the one under scrutiny
from growing, exerting a
sort of external pressure, which may be assumed to come from a quadratic
potential.  A noise term represents atoms in a CZ attaching to other than the proximate island.

To compute
the repulsion, we analyze quantitatively the nucleation of new islands, 
following Ref.~\cite{pimpi2}. If $N$ is the stable island density, $n$ the adatom
density, $D$ the adatom diffusion constant, $\sigma$ the capture coefficient of
an island, and $N_i$ the density of critical nuclei (islands with $i$ atoms),
the nucleation rate $\dot N$ in 2D is \cite{pimpi2}
\begin{equation}
\label{eq:2d}
\dot N = \sigma nN_i\approx Dn^{i+1}.
\end{equation}
In the rightmost expression we have used $\sigma\approx D$ \cite{pimpi2} and the Walton
relation $N_i\approx n^i$ \cite{ven}.
In 2D, $n$ satisfies
\begin{equation}
\dot n = F-\sigma nN\approx 0 \Rightarrow n\approx F/(\sigma N)\approx F\ell^2/D
\label{eq:ndotp}
\end{equation}
where $1/N\approx \ell^{2}=D\tau$ is the squared diffusion length of adatoms before
capture by an island in lifetime $\tau$.
In 1D, Eq.~(\ref{eq:2d}) still holds, but accounting for the properties of a random
walk in 1D makes the capture coefficient $\sigma$ dependent on $\ell$. Indeed, from
\cite{pimpi2,RW2d} we have
\begin{equation}
\label{eq:1d}
\dot N =\sigma_{\rm 1D} nN_i={n\over\tau}N_i(D\tau)^{1/2}\approx {D\over\ell}
nN_{i}\approx {D\over\ell} n^{i+1},\end{equation}
whence $\sigma_{\rm 1D}=D/\ell$. Since $n\propto\ell^2$, regardless of $d$, and
$1/N\approx\ell$ in 1D, Eqs.~(\ref{eq:2d}) and (\ref{eq:1d}) can, for $d$=1,2, be written
\begin{equation}
\label{eq:12d}
\dot N \approx \sigma  n^{i+1}, \;\;\;\;\;\; \sigma=D/\ell^{2-d}.
\end{equation}

Taking the result for $\dot N/\sigma$ in Eq.~(\ref{eq:12d}) as a multiplicity, we find an effective 
entropy $\Sigma=k_B\ln(n^{i+1})$. Identifying $s\equiv\ell^d$ as the ``area'' in $d$=1,2, and recalling
$n \propto \ell^2$ 
\begin{equation}
 (\dot s)_1= K \frac{\partial \left(\Sigma/k_B\right)}{\partial s}=
K~{(\nicefrac{2}{\footnotesize\it d})(i+1)\over s},
\end{equation}
where $K$ is a kinetic coefficient.
Fluctuating repulsion (with strength $B$) from the neighboring CZs yields a second
contribution
$(\dot s)_2=-KBs+\eta$, where $\eta$ arises from
the random component of the external
pressure. We arrive at the Langevin equation
\begin{equation}
\dot s = K \left[(\nicefrac{2}{\footnotesize\it d})(i+1)/ s -B s \right]+ \eta.
\label{eq:sdot}
\end{equation}
\noindent As we show in Ref.~\cite{pimpi1}, the stationary solution to the 
Fokker-Planck equation corresponding to Eq.~(\ref{eq:sdot}) is just
the GWS $P_\beta(s)$, with $\beta \! =\! (\nicefrac{\footnotesize 2}{\it d})(i\! +\! 1)$.

Mulheran and Blackman \cite{muhl,MulGam} proposed the Gamma distribution as an
empirical description of general
Voronoi tessellations, particularly of the CZ size distribution. With unit mean enforced, it has the form \cite{gamGWS} 
\begin{equation}
\Pi_\alpha(x)=\left[ \alpha^{\alpha}/ \Gamma(\alpha)\right]~ x^{\alpha-1}
\exp(-\alpha x).
\label{eq:gam}
\end{equation}
\noindent
No relation was established between the parameter $\alpha$ 
and any
nucleation property, and in accounting for a range of $i$, a single value of $\alpha$ associated with random deposition is used \cite{muhl}. The GWS and Gamma distributions are qualitatively similar, and, for $1\! \le\! \beta \! \le\! 4$, $\alpha$ is roughly $2\beta + \alpha_0$, where $\alpha_0$ is an offset of order one (cf.\ $P_3(s)$ and $\Pi_7(s)$ in Fig.~(\ref{f:WigGam})); the value of $\alpha_0$ depends on what property of the two distributions are equated \cite{Bval}.  However, the slower decay of the $\Pi_\alpha(s)$ leads to considerably greater skewness, with a distinctly greater shift of the peak to smaller $s$.   Like $P_\beta(s)$, $\Pi_\alpha(x)$ approaches a Gaussian for large $\alpha$. Trying to distinguish the two forms in the large-$s$ tail is problematic since the small values lead to large fractional uncertainty.  
Very recently $\Pi_\alpha(s)$ has been used as a tool
for analyzing experimental CZ distributions
\cite{stras,Fanf}. Trial fits of the data with the GWS form are generally at least as good.   Amar {\it et al.}'s popular rate-equation-derived expression for ISD's, noted at the outset, is $f_i(s) \propto s^i\exp(-ia_is^{1/a_i})$, $i\! \ge \! 1$, where $a_i$ is a complicated constant \cite{amfam}.  By construction, it peaks at $s\! = \! 1$.  While not designed for CZ distributions, $f_i(s)$ has been tried as an alternative to $\Pi_\alpha(s)$ for quantum dots, with neither being fully satisfactory \cite{Fanf}. 

Several extensions and challenges present themselves: 1) We hope experimentalists \cite{stras,Fanf} will refit their data for CZ distributions with the GWS. When Voronoi tessellation is used more generally, e.g.\ in studying biological systems \cite{lipid}, the resulting size histograms should be analyzed using the GWS and the alternatives.  2) Fig.~\ref{f:isdcz}a shows that when there is a wide range of island sizes, using a strict Voronoi construction rather than physically more appropriate edge cells \cite{evans1,evans2} leads to a narrower distribution (hence a higher deduced $\beta$).  E.g., in fitting numerical data for the irreversible point-island nucleation ($i$=1) in Fig.~3b of Ref.~\cite{evans1}, we find much better agreement with $P_4(s)$ than the expected $P_2(s)$.  (Other subtleties complicate this case; compact islands occur only at very low coverage ($\leq$0.01) \cite{evans1}.) 3) Our predictions should be tested in cases with large $i$. 4) For $d\! >$2, random walks are not recurrent, with the upshot that $\beta$=$i$+1 as for $d$=2. This can be tested with growth simulations in $d$=3  and $d$=4 \cite{amars}. 5) In $d$=1 it is straightforward to find the gap distribution between islands, as done in Ref.~\onlinecite {bm1996}.  The CZ distribution is a self-convolution of the gap distribution \cite {bm1996}. In Fig.~\ref{f:bm} we saw $\beta$=4; the gap distribution, then, should be well described by $P_{3/2}(s)$ (viz.\ $P_4(s) \approx 2\int_0^{2s}P_{3/2}(s^\prime)P_{3/2}(2s\! -\!s^\prime)\, ds^\prime$).  The inset of Fig.~\ref{f:bm} corroborates this.  
Generalization to $d$=2 is unclear.

In summary, as for spacings between parked cars \cite{Abul}, the Wigner surmise provides a simple, universal expression that accounts better for data than more complicated expressions developed over years of investigation.  For our problem of the capture
zone distribution in island nucleation, the exponent $\beta$ of the generalized Wigner surmise $P_\beta(s)$ provides information about the size $i$ of the critical nucleus and reflects the dimensionality $d$.  Our phenomenological argument provides insight into the physical origin of this behavior.  Both features are significant advances beyond previous empirical analytic
descriptions of the CZ size distribution (notably $\Pi_\alpha(x)$).  The connection to universal properties 
of fluctuations enhances the interest and importance of studies of CZ distributions and suggests many avenues for further investigations.

\section*{Acknowledgments}

Work at the University of Maryland was supported by 
the NSF-MRSEC, Grant DMR 05-20471. Visits to Maryland by A.P. were supported by a CNRS Travel Grant, and T.L.E. was partially supported by DOE CMSN grant DEFG0205ER46227.


\begin{thebibliography}{99}
\bibitem{notlog} The form of the stationary distribution does not depend sensitively on the form of the long-range repulsion [\onlinecite{Abul}].
\end{thebibliography}
\end{document}